\documentclass[twocolumn, pre]{revtex4}

\usepackage{amssymb}
\usepackage{amsmath}
\usepackage{graphics}
\usepackage{graphicx}

\newcommand{\be}{\begin{equation}}
\newcommand{\ee}{\end{equation}}

\begin{document}

\title{Anomalously large capacitance of a plane capacitor with a two-dimensional electron gas}

\date{\today}
\author{Brian Skinner}
\author{B. I. Shklovskii}
\affiliation{Fine Theoretical Physics Institute, University of Minnesota, Minneapolis, Minnesota 55455}

\begin{abstract}

In electronic devices where a two-dimensional electron gas (2DEG) comprises one or both sides of a plane capacitor, the resulting capacitance $C$ can be larger than the ``geometric capacitance" $C_g$ determined by the physical separation $d$ between electrodes.  This larger capacitance is known to result from the Coulomb correlations between individual electrons within the low density 2DEG, which lead to a negative thermodynamic density of states.  Experiments on such systems generally operate in the regime where the average spacing between electrons $n^{-1/2}$ in the 2DEG is smaller than $d$, and these experiments observe $C > C_g$ by only a few percent.  A recent experiment \cite{Ashoori}, however, has observed $C$ larger than $C_g$ by almost 40\% while operating in the regime $nd^2 \ll 1$.  In this paper we argue that at $nd^2 \ll 1$ correlations between the electronic charge of opposite electrodes become important. We develop a theory of the capacitance for the full range of $nd^2$.  We show that, in the absence of disorder, the capacitance can be $4d/a$ times larger than the geometric value, where $a \ll d$ is the electron Bohr radius.  Our results compare favorably with the experiment of Ref.\ \onlinecite{Ashoori} without the use of adjustable parameters.

\end{abstract} \maketitle

\section{Introduction}

In a standard parallel-plate capacitor, the capacitance $C$ is
equal to the ``geometric capacitance" $C_g = \varepsilon S/4 \pi
d$ (in Gaussian units), where $\varepsilon$ is the dielectric
constant of the medium separating the two plates, $S$ is the area
of each plate, and $d$ is the separation between them.  The
expression $C = C_g$ is correct when both electrodes are made from
a ``perfect" metal, which by definition screens electric field
with a vanishing screening radius, so that the charge of a given
electrode is located exactly on the electrode surface and the
electric field from the opposite electrode does not penetrate into
the metal.  If one of the electrodes is made from a material with
finite (positive) Debye screening radius $R_D$ (for example, a
doped bulk semiconductor), then the imperfect charge screening at
this electrode allows the electric field to penetrate a distance
$R_D$ into the electrode and the capacitance decreases.  If one
describes the capacitance by the effective capacitor thickness
$d^* = \varepsilon S/4 \pi C$, then the effect of positive
screening radius is to increase the effective capacitor thickness
from $d^* = d$ to $d^* = d + R_D$.

On the other hand, capacitors with $d^* < d$, or in other words
with effective three-dimensional electrode screening radius $R_D <
0$, are known in semiconductor
physics~\cite{BLES1981,Luryi,KRAV1990,Eis1992,Eis-long,Sivan,Jiang,Yacoby,Allison,SE,Efros92,Pikus92,Pikus93,Shi,Fogler,Efros08,Kopp}.
Examples include Si MOSFETs and gated GaAs-AlGaAs
heterostructures, where one electrode consists of a clean, low
density, two-dimensional electron gas
(2DEG)~\cite{KRAV1990,Eis1992,Eis-long,Sivan,Jiang,Yacoby,Allison}.
In these devices, the total capacitance can be written as \be
\frac{1}{C} = \frac{1}{C_g} + \frac{d\mu/dn}{S e^2}, \label{eq:1C}
\ee where $n$ is the electron area density, $\mu$ is the chemical
potential of the 2DEG, and $e$ is the elementary charge. In terms
of the effective thickness $d^*$, Eq.\ (\ref{eq:1C}), implies 
\be
d^* = d + r_D/2, 
\ee 
where $r_D = \varepsilon d\mu/dn /(2 \pi e^2)$ is the Debye screening radius of the 2DEG. Capacitance larger than the geometric value, or $r_D < 0$, is possible when the thermodynamic density of states $dn/d\mu$ of the 2DEG is negative (or, equivalently, when the compressibility $(n^2 d\mu/dn)^{-1} < 0$).

In the limit of low electron density, such that the average
distance $n^{-1/2}$ is much larger than the effective Bohr radius
$a = \varepsilon \hbar^2/m e^2$ of the electrons, or in other
words the dimensionless parameter $r_s = (\pi n a^2)^{-1/2} \gg 1$, a
2DEG is a classical system whose physics is dominated by the
Coulomb interaction between electrons. This interaction leads to a
Wigner crystal-like strongly-correlated liquid state with negative
chemical potential $\mu \simeq -2.9 e^2 n^{1/2}/\varepsilon$.  The
corresponding Debye screening radius  $r_D = - 0.23 n^{-1/2}$
produces a negative correction~\cite{BLES1981,foot} to $d^*$:
\be
d^* \simeq d - 0.12 n^{-1/2}, \hspace{5mm} (n^{1/2}d \gg 1).
\label{eq:dhigh}
\ee

What happens to $d^*$ when $n^{-1/2} \gg d$?  This is the main question addressed in this paper.  Thermodynamic stability criteria ensure that the capacitance cannot be negative \cite{LL8}, so Eq.\ (\ref{eq:dhigh}) must not apply at such low densities.  In Sec.\ \ref{sec:metal} of this paper we find the function $d^*(d, n) = d \cdot f(n^{1/2}d)$, valid over the whole range of $nd^2$.  The dimensionless function $f(x)$ is shown in Fig.\ \ref{fig:dstar-metal}. We show that in the limit $nd^2 \ll 1$ the effective thickness $d^*$ becomes very small:
\be
d^* = 2.7 d (n d^2)^{1/2}, \hspace{5mm} (n^{1/2}d \ll 1).
\label{eq:dlow}
\ee
This dramatic capacitance growth is due to the coupling of each
electron in the 2DEG to its image charge in the metal electrode.
At low density, compact electron-image dipoles are separated from
each other by a distance much larger than their dipole arm.
These dipoles interact weakly with each other, providing only a
small resistance to capacitor charging.

\begin{figure}[htb]
\centering
\includegraphics[width=0.45 \textwidth]{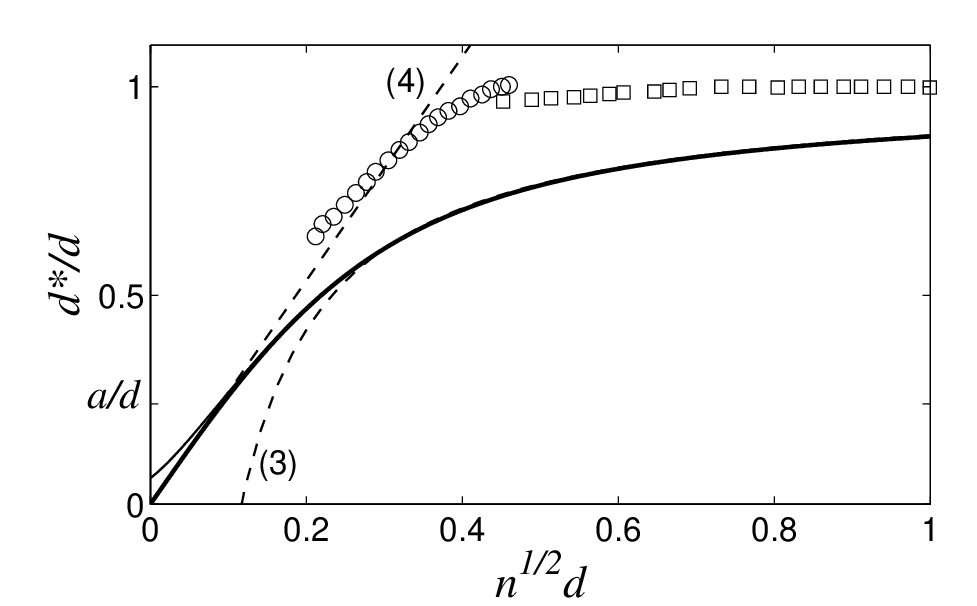}
\caption{The effective thickness $d^*/d$ of a capacitor made from
a 2DEG and a metal electrode as a function of the dimensionless
parameter $n^{1/2}d$.  Open squares and circles correspond to data
from Ref.\ \onlinecite{Ashoori} for devices 1 and 2, respectively; in each case the geometric capacitance $C_g$ was assumed to be equal to the largest recorded value of the capacitance.  The thick solid curve, which contains no adjustable parameters, is the prediction of this paper for a classical 2DEG.  At relatively large density, there is a small downward correction to $d^*$, described by Eq.\ (\ref{eq:dhigh}).  At $n^{1/2}d \ll 1$, $d^*$ is described by Eq.\ (\ref{eq:dlow}).  The thin solid line schematically shows the role of quantum mechanical motion for finite $a$, which at $d^* \approx a$ produces a deviation from Eq.\ (\ref{eq:dlow}) and leads to a saturation of the capacitance at $d^* = a/4$.  To draw it we used $a = d/4$.} \label{fig:Ashooricompare}
\end{figure}

Until recently, only relatively small corrections to the geometrical capacitor thickness $d$ were observed experimentally \cite{KRAV1990,Eis1992,Eis-long,Sivan,Jiang,Yacoby,Allison}.  The most recent paper on this subject~\cite{Ashoori}, however, claims a much larger correction $(d-d^*)/d \sim 0.4$ for a YBa$_2$Cu$_3$O$_7$/LaAlO$_3$/SrTiO$_3$ (YBCO/LAO/STO) capacitor with a 2DEG at the LAO/STO interface separated by $d = 4$ nm of LAO
insulator from the metallic YBCO gate.  Theoretical estimates show that $a \sim 1$ nm is by far the smallest length scale in the problem, so that as a zero-order approximation one can use a classical description of the 2DEG.  This measurement calls for a comparison of our function $d^*(d, n)$ with the experimental data of Ref.\ \onlinecite{Ashoori}.  Such a comparison is shown in Fig.\ \ref{fig:Ashooricompare} and looks quite good without the use of adjustable parameters.  We return to a more detailed  discussion of this comparison in Sec.\ \ref{sec:discussion}.

In Sec.\ \ref{sec:2degs} we study the case where both sides of the
capacitor are made from 2DEGs with equal electron density $n$.
Such a capacitor in principle can be realized in devices with two
parallel quantum wells with tunable concentrations of electrons
\cite{Eis-long}, but we do not know of any published results.
The capacitance of such devices was addressed theoretically in
Ref.\ \onlinecite{Kopp}. In the limit $nd^2 \gg 1$, the authors of Ref.\ \onlinecite{Kopp} arrive at a small correction to $d$ which is twice larger than for that of one 2DEG: $d^* = d + r_D$. In the classical limit $r_s \gg 1$, this gives $d^* = d - 0.23 n^{-1/2}$.  The authors of Ref.\ \onlinecite{Kopp} assume that this equation remains valid even at
$nd^2 \ll 1$, which leads them to the prediction that $C$ diverges
and becomes negative at a finite value of $nd^2$.  As the authors themselves recognized, however, their assumption ignores correlations between the two 2DEGs.  In  Sec.\ \ref{sec:2degs} we account for these correlations and demonstrate that they dramatically alter the results of Ref.\ \onlinecite{Kopp} for $d^*$.  Namely, $d^*$ vanishes and $C$ diverges only in the limit $nd^2 \rightarrow 0$, as in the case of a single 2DEG.  The transition from large to small $nd^2$ for two 2DEGs is described by the equation $d^* = d \cdot f(n^{1/2}d/2)$, where $f(x)$ is the same function as for the single 2DEG case. Also similar to the single 2DEG case, the diverging capacitance can be explained by
strong correlations between the Wigner crystals of the two 2DEGS,
so that an electron transferred from one 2DEG to the other is
still bound to its image charge (the hole left behind in the
opposite electrode, see Fig.\ \ref{fig:2DEGs}). Thus, at $nd^2 \ll
1$ only the weak dipole-dipole repulsion between two electron-hole
dipoles is responsible for resistance to charging of the 2DEGs capacitor.

Of course, the divergence of the capacitance at $nd^2 \rightarrow 0$
takes place only at simultaneously vanishing Bohr radius $a$, temperature, disorder, and, in the case of two 2DEGs, probability of tunneling through the insulator.  At some strength these factors destroy the Coulomb correlations between electrons at a particular value of $nd^2$ and truncate the capacitance growth, so that at $nd^2 \rightarrow 0$ the capacitance remains finite.  Such a behavior is shown schematically by the thin line in Fig.\ \ref{fig:Ashooricompare}, which assumes vanishing temperature and disorder but finite $a = d/4$.  We see that the capacitance can grow as much as 16 times from the geometrical value.

Usually, disorder is so severe that it closes the window of $nd^2$ in which $d^*$ is substantially smaller than $d$.  Nonetheless, the experiment of Ref.\ \onlinecite{Ashoori} shows a large correction, so that apparently such a window is open.  In Sec.\ \ref{sec:discussion} we discuss this experiment in greater detail in an attempt to understand why it represents a special case where large capacitance can be observed.  We also discuss the effects of the quantum kinetic energy of electrons in the 2DEG, and show that in the absence of disorder it  provides an upper limit for the capacitance at $d^* = a/4$.

We note that this paper represents an extension of an approach we have previously used to study large capacitance at the interface between a metal and an ionic conductor (an ion-conducting glass \cite{us-PRL, us-longer} or an ionic liquid \cite{Loth}).  Such interfaces block both ionic and electronic current, thereby forming a capacitor even in the absence of an insulating layer.  The binding of discrete ions to their image charges in the metal results in a weaker, dipole-dipole repulsion between counterions and therefore in large capacitance of the interface.  The resulting effective thickness of the capacitor can, surprisingly, be even smaller than the ion radius.  This paper describes a similar effect for systems where the countercharge consists of a 2DEG separated from the metal by an insulator.  Over a certain range of the electron density $a^2/d^4 \ll n \ll 1/a^2$, the capacitance in such systems is dominated by the strong Coulomb interactions between discrete charges and can therefore be described using a classical analysis similar to that of Refs.\ \onlinecite{Loth, us-PRL, us-longer}.  A very brief report about the first part of this work was published in our recent preprint \cite{Loth}.

Our general approach to calculating the capacitance in the sections below is as follows.  We first describe the total electrostatic energy $U(n)$ associated with the ground state configuration of $n$ electrons per unit area.  If the two sides of the capacitor are coupled through a voltage source with voltage $V$, then the value of the charge $Q$ of the capacitor is that which minimizes the total energy $U - QV$, where the term $-QV$ represents the work done by the voltage source relative to the situation $V = 0$.  Using the (zero-temperature) equilibrium condition $d(U - QV)/dQ = 0$ along with $dQ = e S dn$ gives
\be 
V = \frac{d U}{d Q} = \frac{1}{eS} \frac{d U}{dn}.
\label{eq:Vdef}
\ee
The differential capacitance (or ``charge susceptibility") of the system $C = (dV/dQ)^{-1}$ can therefore be written
\be 
C = e^2 S^2  \left( \frac{d^2 U}{d n^2} \right)^{-1}. \label{eq:Cdef} 
\ee
We can solve for the capacitance as a function of voltage, $C(V)$, by combining this relation with Eq.\ (\ref{eq:Vdef}).  Finally, the effective capacitor thickness $d^*$ is also defined by the total energy $U$ as
\be 
d^* = \frac{\varepsilon}{4 \pi e^2 S} \frac{d^2 U}{d n^2}. \label{eq:dstardef} 
\ee 
In this way a description of the total energy is sufficient to determine the capacitance, and it is not necessary to invoke the Poisson equation or to make mean-field approximations of the electric potential.

\section{Capacitor with a classical 2DEG and a metal electrodes} \label{sec:metal}

In this section we describe a 2DEG with area density $n$ separated from a perfect metal electrode by an insulator of thickness $d$.  This can be, for example, a Si MOSFET or a gated GaAs-AlGaAs
heterostructure, where a 2DEG is created at the
semiconductor-insulator interface and connected to one terminal of
a voltage source by ohmic contacts.  The metal electrode is
connected to the opposite terminal of the voltage source.  For
simplicity, we assume that the dielectric constant $\varepsilon$
is uniform everywhere.  We treat the 2DEG in the classical limit $na^2 \ll 1$.

\begin{figure}[htb]
\centering
\includegraphics[width=0.35 \textwidth]{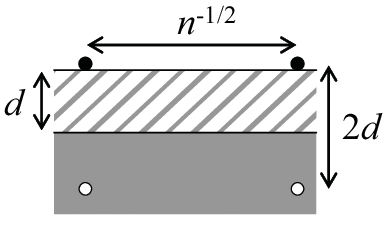}
\caption{Two neighboring electrons (black circles) in a 2DEG
formed at the semiconductor-insulator interface.  The electrons
are separated from a metal electrode (solid area) by an insulator
of thickness $d$ (hatched area).  The electrons form positive
image charges in the metal (white circles).} \label{fig:insulator}
\end{figure}

In the ground state for low electron density $n$, the repulsion
between electrons within the 2DEG causes them to form a
strongly-correlated liquid, reminiscent of a two-dimensional
Wigner crystal, in which electrons are separated from their
nearest neighbors by a distance $\sim n^{-1/2}$.  Each electron,
of charge $-e$, also induces an image charge, $+e$, in the metal
surface, which is effectively located a distance $2d$ from the
2DEG.  This situation is shown schematically in Fig.\
\ref{fig:insulator}.  We suppose that the 2DEG is connected to
some voltage source which maintains a difference in electric
potential $V$ between the 2DEG and the metal electrode.  The
charge $Q$ of the capacitor is defined as the amount of charge
that has moved through the voltage source relative to the state $V
= 0$.

A given electron within the 2DEG and its image charge effectively
form an electron-image dipole, with dipole moment $2ed$, which
repels an adjacent electron.  In the limit $n^{-1/2} \gg d$, we
can use the point-dipole approximation and the repulsive energy
between two adjacent electrons is $2 e^2 d^2 n^{3/2}/\varepsilon$.
More generally, we can write the total electrostatic energy of the
$nS$ electrons in the 2DEG by first calculating the electrostatic
potential $\phi_0$ experienced by each electron relative to
infinity.  Specifically, for an electron located at the origin
\be
\phi_0 = \frac{e}{2\varepsilon d} - \sum_{\{i,j\} \neq \{0,0\}}
\frac{e}{\varepsilon} \left( \frac{1}{r_{i,j}} -
\frac{1}{\sqrt{r_{i,j}^2 + (2d)^2}} \right),
\label{eq:phi0}
\ee
where the indices $i,j$ label the set of electron locations and
$r_{i,j}$ is the distance between the electron $\{i,j\}$ and the
origin. The term outside the sum in Eq.\ (\ref{eq:phi0}) indicates
the potential contributed by the electron's own image charge, which is added in place of the self-interaction term $\{i,j\} = \{0,0\}$. $\phi_0$ can be estimated by assuming that the electrons occupy a regular square lattice with lattice constant $n^{-1/2}$, in which case \be
\phi_0 = \frac{e}{2\varepsilon d} - \frac{e n^{1/2}}{\varepsilon} \cdot g(n^{1/2}d) ,
\label{eq:phi02}
\ee
where $g(x)$ is a dimensionless function
\be
g(x) = 4 \sum_{i = 1}^\infty \sum_{j = 0}^\infty \left( \frac{1}{\sqrt{i^2 + j^2}} - \frac{1}{\sqrt{i^2 + j^2 + 4 x^2}} \right).
\label{eq:g}
\ee
The sum in Eq.\ (\ref{eq:g}) is convergent for all $x$.  We note that
while the true lowest energy configuration for the electrons is to
occupy a triangular lattice, the energy per unit area of a square
lattice of dipoles differs from that of a triangular lattice by less than $2\%$ \cite{Topping}, so for computational simplicity we use a
square lattice for all calculations.

The total energy of the configuration of electrons is
\be
U = -\frac12 e S n \phi_0 = -\frac{e^2 S}{4 \varepsilon d} n +
\frac{e^2 S}{2 \varepsilon}  n^{3/2} g(n^{1/2}d).
\label{eq:U}
\ee
Combining Eqs.\ (\ref{eq:U}) and (\ref{eq:dstardef}) gives 
\be 
\frac{d^*}{d} = \frac{1}{32 \pi} \left[ \frac{3 g(x)}{x} +5 g'(x) + x g''(x) \right] \equiv f(x), 
\label{eq:dstarexact} 
\ee 
where $x = n^{1/2}d$.

\begin{figure}[htb]
\centering
\includegraphics[width=0.45 \textwidth]{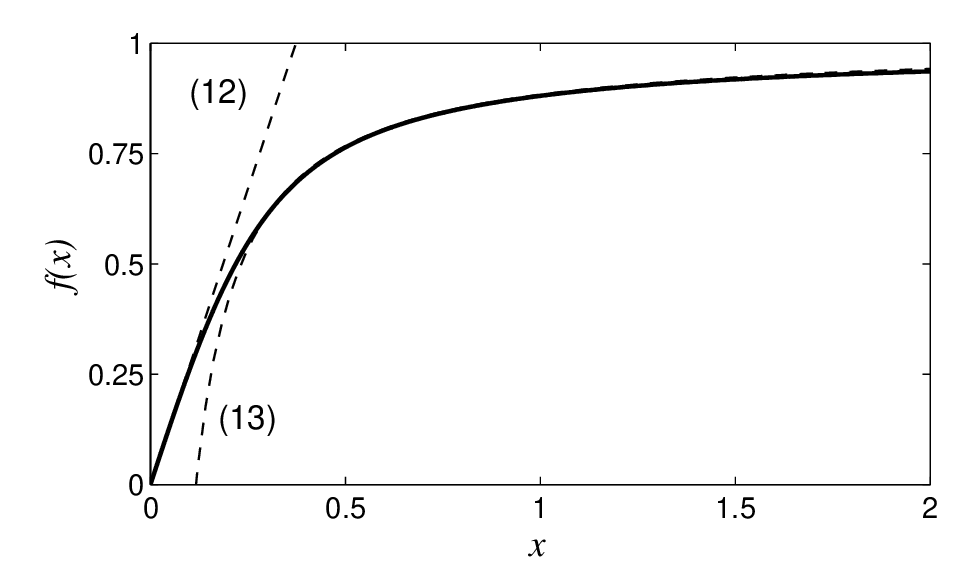}
\caption{The dimensionless function $f(x)$ defined in Eq.\
(\ref{eq:dstarexact}).  The function gives the effective thickness
$d^*/d$ of a capacitor composed of a 2DEG separated from a metal
surface by an insulator of thickness $d$ as a function of
the dimensionless electron density $x = n^{1/2} d$.  The left side
of the plot corresponds to a very sparse 2DEG, where the electrons
can be thought to form an array of discrete electron-image dipoles
and $d^*$ is described by Eq.\ (\ref{eq:dstar1}).  The right side
corresponds to a relatively dense packing of electrons, where the
electrons form an almost uniform layer of charge and $d^*$ is
described by Eq.\ (\ref{eq:dstar2}).} \label{fig:dstar-metal}
\end{figure}

The dimensionless function $f(x)$ is plotted in Fig.\
\ref{fig:dstar-metal}.  At $x \ll 1$, one can expand the summand
in Eq.\ (\ref{eq:g}) to lowest order in $x$ and arrive at the
point-dipole approximation for the interaction among
electron-image pairs, which after summation gives $g(x) \simeq 18
x^2$.  The resulting effective capacitor thickness approaches zero
linearly with $x$, and the function $f(x)$ is described by 
\be
f(x) \simeq 2.7 x, \hspace{5mm} (x \ll 1), 
\label{eq:dstar1} 
\ee
which is equivalent to Eq.\ (\ref{eq:dlow}).  This vanishing of
$d^*$ at $x \rightarrow 0$ implies that the capacitance diverges
when the electron gas is very sparse.  Such diverging capacitance
is the result of a vanishing dipole-dipole repulsion between
adjacent electron-image pairs as the electron density goes to
zero.

At large electron density $x \gg 1$, the effective thickness $d^*$
approaches the geometric thickness $d$.  In other words, $C$
approaches $C_g$.  In the region $x \gg 1$, the difference between
$d^*$ and $d$ can be viewed as a small correction associated with
a finite negative screening radius of the 2DEG.  In this case
$f(x)$ approaches 
\be 
f(x) = 1 - 0.12/x, \hspace{5mm} (x \gg 1),
\label{eq:dstar2} 
\ee 
which is equivalent to Eq.\
(\ref{eq:dhigh}). It is only at much larger density $n \gtrsim
1/a^2$, that quantum effects cause the 2DEG screening radius $r_D$ to become positive, so that $d^*$ becomes larger than $d$.

\begin{figure}[htb]
\centering
\includegraphics[width=0.45 \textwidth]{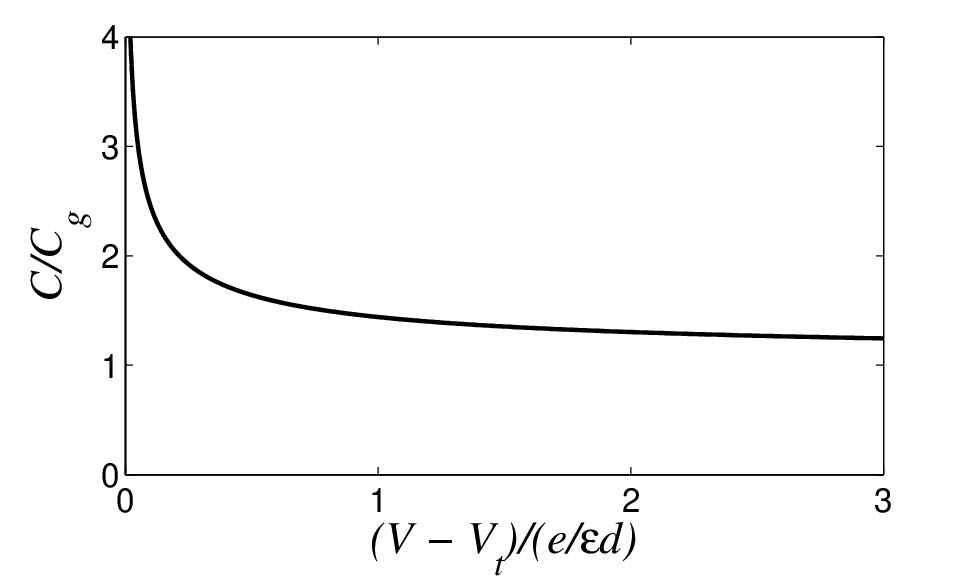}
\caption{The zero-temperature capacitance of a capacitor made from
a 2DEG parallel to a perfect metal electrode, plotted as a
function of voltage.} \label{fig:C-V}
\end{figure}

We can also derive a relation between the voltage of the capacitor
and the electron density $n$ by using Eq.\ (\ref{eq:Vdef}).  If
we define $V_t$ to be the ``threshold voltage" at which the
electron gas is completely depleted ($n = 0$), then the derivative
of Eq.\ (\ref{eq:U}) implies that
\be
V - V_t = \frac{e}{4 \varepsilon d} x \left[ 3 g(x) + x g'(x) \right]. \label{eq:Vn}
\ee
Combining the results of Eqs.\ (\ref{eq:dstarexact}) and
(\ref{eq:Vn}) allows us to create a plot of the capacitance as a
function of $V - V_t$.  The result is shown in Fig.\ \ref{fig:C-V}, with the capacitance and voltage plotted in the dimensionless forms $C/C_g$ and $V/(e/\varepsilon d)$, respectively.  At small voltages $0 < V - V_t \ll e/\varepsilon d$, the capacitance diverges as $C \propto (V - V_t)^{-1/3}$.  At large voltages $V - V_t \gg e/\varepsilon d$,
the capacitance approaches its geometric value.

\section{Capacitor made from two 2DEGs} \label{sec:2degs}

\begin{figure}[htb]
\centering
\includegraphics[width=0.45 \textwidth]{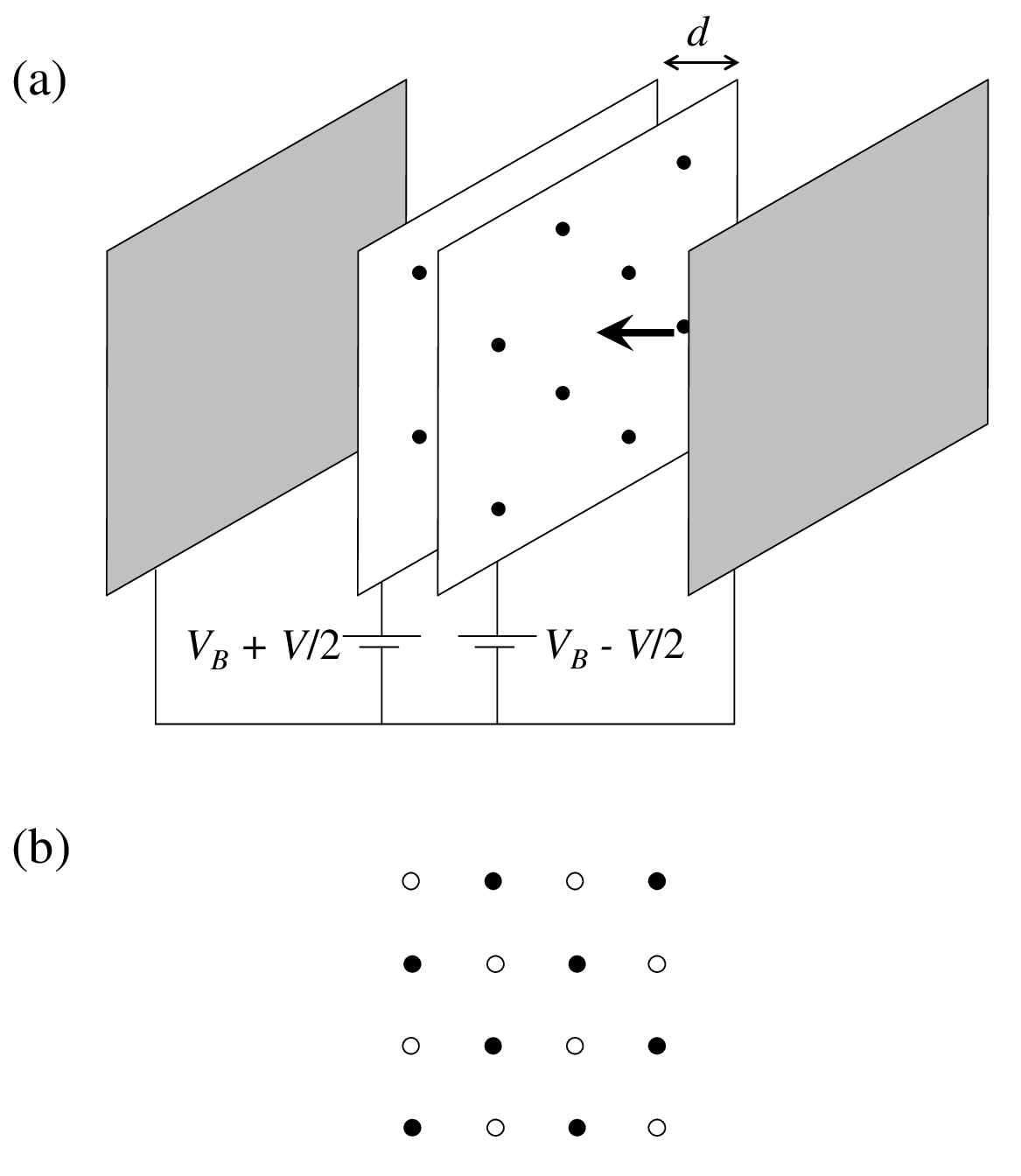}
\caption{(a) A schematic picture of a capacitor made from two parallel, depleted electron gases (black circles on top of white squares).  A large bias voltage $V_B$ is applied from either side of the 2DEGs by metallic gates (shaded squares) in order to deplete the 2DEGs to some small density $n$.  An additional small voltage $V$ is applied between the two 2DEGs and determines the capacitance.  Insulating layers separating the metallic gates from the 2DEGs and the 2DEGs from each other are not shown.
(b) A frontal view of the 2DEGs, as indicated by the thick arrow in (a).  On the upper 2DEG, electrons (black circles) form a lattice with lattice constant $\sim n^{-1/2}$.  Electrons on the lower 2DEG (white circles) also form a lattice, but this lattice is offset from the other so that electrons in the two 2DEGs minimize their Coulomb interaction
energy.} \label{fig:2DEGs}
\end{figure}

One can also imagine a capacitor where 2DEGs comprise \emph{both}
electrodes, as was treated theoretically in Ref.\ \onlinecite{Kopp}.  Such a situation is possible in devices with two parallel quantum wells with tunable concentrations and separate contacts.  This can be, for example, a GaAs-AlGaAs-GaAs heterostructure, where 2DEGs are formed at both heterojunctions.  The densities of the two 2DEGs can be tuned by applying a large bias voltage $V_B$ above each of them.  If a small additional voltage $V$ is applied between the two 2DEGs, then the response to this small voltage can be used to determine the capacitance of the two-2DEG system.  This setup is shown schematically in Fig.\ \ref{fig:2DEGs}(a).

In this section we consider the case of two identical 2DEGs
oriented parallel to each other and separated by a distance $d$.
They are connected to opposite terminals of a voltage source
maintained at a particular voltage $V$.  We assume that at $V = 0$
both 2DEGs have the same density $n$ of electrons and that charge neutrality is maintained by a uniform plane of surface charge with density $+en$ that coincides with the plane of each 2DEG.  We also assume, for simplicity, that the dielectric constant $\varepsilon$ is uniform everywhere.

At zero temperature, the electron positions are strongly
correlated, with electrons in a 2DEG seeking to maximize their
separation both from each other and from electrons in the opposite
plane.  As a result, at $V=0$ electrons form interlocking lattices
of electron positions on the two electrodes, as shown
schematically in Fig.\ \ref{fig:2DEGs}(b).  The exact configuration of the two lattices can take on one of three arrangements, depending on the value of the parameter $nd^2$ \cite{Goldoni}.  However, the energy of these different lattice types differs by only a few percent, so for illustrative purposes we have shown the simplest case of two interlocking square lattices.

In order to give the two-2DEG capacitor a finite charge $Q$, some
number of electrons must be transferred from one 2DEG to the
other.  For one electron, this process requires a finite amount of energy $\Delta u$, associated with creating a defect in the two lattices.  While a careful calculation of $\Delta u$ is not a major goal of this paper, we make an estimate of its value at the end of this section.

When the voltage applied between the two 2DEGs is smaller than
$\Delta u/e$, no charge transfer is possible and the capacitance
$C = 0$.  At $V > \Delta u/e$, some finite area density of
electrons $\delta n$ is transferred from one 2DEG to the other and
the corresponding capacitor charge is $Q = e S\delta n$.  These
``excess electrons" also repel each other, and they seek to
maximize their distance from each other by forming a Wigner
crystal-like lattice of defects in the ground state
``checkerboard" of electrons.  At low temperature, excess
electrons remain coupled to the ``holes" they leave behind in the
opposite 2DEG and therefore they repel each other by a
dipole-dipole repulsion.  The form of this repulsion is identical
to that of the previous section, where the image charge was formed
in the metal electrode, except that in the present case the dipole
arm is $d$ rather than $2d$ and there is an overall factor $2$
associated with the presence of a repulsive force at both positive
and negative sides of the dipole.  This similarity allows us to
use previous results in writing the total electrostatic energy $U$
of the system relative to the ground state.  Namely,
\be
U = S \delta n \Delta u + \frac{e^2 S}{\varepsilon} (\delta n)^{3/2} \cdot g \left( (\delta n)^{1/2} d/2 \right),
\ee
where $g(x)$ is the same function defined in Eq.\ (\ref{eq:g}).

The density of excess electrons $\delta n$ can be related to the
voltage $V$ by $V = dU/dQ = dU/d(e S \delta n)$.  If we define $y
= (\delta n)^{1/2} d$, then this relation gives
\be V - \frac{\Delta u}{e} = \frac{e}{2 \varepsilon d} \cdot \frac{y}{2} \left[ 3 g(y/2) + \frac{y}{2} g'(y/2) \right]. \label{eq:Vny}
\ee
As in Eq.\ (\ref{eq:dstardef}), the corresponding effective
thickness is $d^* = \varepsilon/4 \pi e^2 S \cdot d^2 U/d(\delta
n)^2$, which gives
\be
\frac{d^*}{d} = f(y/2),
\label{eq:dstarexact2}
\ee
where $f(x)$, plotted in Fig.\ \ref{fig:dstar-metal}, is the same function as in Eq.\ (\ref{eq:dstarexact}).

Eq.\ (\ref{eq:dstarexact2}) is correct when the applied voltage
$V$ is low enough in absolute value that neither 2DEG is depleted.
At some critical voltage $V_c$, however, no additional charge
transfer is possible between the two 2DEGs and the capacitance
collapses.  The value of $V_c$ can be estimated by setting $\delta n =
n$, which corresponds to substituting $x = n^{1/2} d$ for $y$ in
Eq.\ (\ref{eq:Vny}).

\begin{figure}[htb]
\centering
\includegraphics[width=0.4 \textwidth]{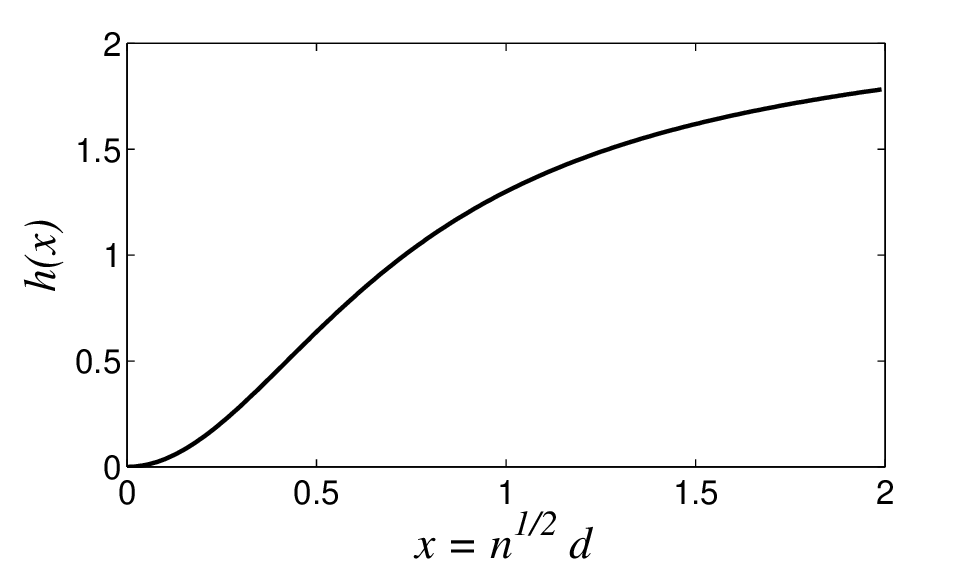}
\caption{The dimensionless function $h(x)$, as defined in Eq.\
(\ref{eq:Deltau}).  This function determines the energy required
to move a single electron from one 2DEG to the other at zero
temperature.} \label{fig:hx}
\end{figure}

We now comment on the threshold energy $\Delta u$ required to transfer a single electron from one 2DEG to the other at zero voltage.  A rough estimate of $\Delta u$ can be made by imagining that an electron is transferred to the site directly across from it in the opposite 2DEG and that all other electrons remain in their ground state positions.  In this case $\Delta u$ can be evaluated as
\be
\Delta u = \frac{e^2 n^{1/2}}{\varepsilon} h(n^{1/2} d),
\label{eq:Deltau}
\ee
where $h(x)$ is a dimensionless function
\be
h(x) = 4 \sqrt{2} \sum_{i = 1}^{\infty} \sum_{j = 0}^{\infty}
\left[ \frac{(-1)^{i+j+1}}{\sqrt{i^2 + j^2}} -
\frac{(-1)^{i+j+1}}{\sqrt{i^2 + j^2 + 2 x^2}} \right].
\label{eq:hx}
\ee
Here, for simplicity, we have again assumed a square lattice of electron positions, as in Fig.\ \ref{fig:2DEGs}.  The function $h(x)$, which is convergent for all $x$, is plotted in Fig.\ \ref{fig:hx}.  The estimate of Eqs.\ (\ref{eq:Deltau}) and (\ref{eq:hx}) assumes that the lattice of electron positions is undisturbed by the transfer of one electron, which is valid when the 2DEGs are relatively sparse, $x \ll 1$, which is the major focus of this paper.  At larger $x$ the lattice is more easily deformed and Eq.\ (\ref{eq:Deltau}) represents an upper bound for $\Delta u$.

\section{Discussion} \label{sec:discussion}

In this section we discuss the truncation of the capacitance divergence of a classical 2DEG, concentrating primarily on the case of a single 2DEG parallel to a metal electrode.  So far we have dealt only with a clean, classical 2DEG.  Of course, at zero temperature and in the absence of disorder the capacitance cannot diverge without bound in the limit $n d^2 \rightarrow 0$ because of the effects of quantum kinetic energy. This conclusion can be reached by considering that each electron within the 2DEG sits in a potential well created by neighboring electron-image dipoles.  If this potential $w$ is expanded to second order in the electron's displacement $\rho$ from the potential well minimum, then we get $w(\rho) = \gamma e^2 d^2 n^{5/2}/\varepsilon \cdot \rho^2 + e \phi_0$, where $\gamma \approx 54$ is a numerical coefficient and $\phi_0 \approx 18 e d^2 n^{3/2}/\varepsilon$ is the electrostatic potential described by Eq.\ (\ref{eq:phi02}).  This potential $w(\rho)$ is that of a two-dimensional harmonic oscillator, and therefore it has a ground state energy $\hbar \omega + e \phi_0$, where $\omega$ is defined so that $w(\rho) = m \omega^2 \rho^2/2 + e \phi_0$.  When $\hbar \omega \ll e \phi_0$, the 2DEG retains its strongly-correlated, Wigner crystal-like structure and the contribution of the quantum kinetic energy to the capacitance is small.  At $n^{1/2} d \lesssim 0.2 a/d$, however, the quantum kinetic energy becomes larger than $e \phi_0$ and the 2DEG loses its electrostatic correlations.  This point corresponds to $d^* \approx a$.  At vanishingly small values of $n^{1/2} d$, the energy of the 2DEG is that of a noninteracting fermion gas, which produces $d^* = a/4$.  For the experiments of Ref.\ \onlinecite{Ashoori}, $d= 4$ nm and $a \approx 1$ nm, so that apparently in this system a capacitance $C \approx 16 C_g$ is possible (see Fig.\ \ref{fig:Ashooricompare}).

The analysis above has also ignored disorder, which can truncate the divergence of the capacitance by destroying dipole-dipole correlations.  The presence of disorder modulates the density of the correlated dipole liquid with some characteristic amplitude $\delta n_d$.  At small enough average density $n$, the overall electron density becomes smaller than $\delta n_d$. This means that screening becomes nonlinear~\cite{SE} and multiple pores open in the 2DEG.  Electric field lines starting at the metal gate electrode leak through these pores. As a result, the Debye screening radius $r_D$ changes its sign \cite{Eis1992,Eis-long,Sivan,Jiang,Yacoby,Allison,SE,Efros92,Pikus92,Pikus93,Shi,Fogler}
from negative to positive at some $n = n_m$. At smaller $n$ the effective thickness $d^*$ grows sharply. In a 2DEG with moderate mobility and large distance to the gate, $n_m d^2
\gg 1$ and the minimum of $d^*/d$ is very shallow. For example, for the classical 2D hole gas (with large $r_s$) in GaAs/GaAlAs heterojunctions studied in Ref.~\cite{Allison}, $n_{m}d^2 = 5 $, while as we see in Fig. \ref{fig:dstar-metal} the crossover between the $n d^2 \gg 1$ and $n d^2 \ll 1$ asymptotic dependencies happens only around $n^{1/2}d \sim 0.25$ or $n d^2 \sim 0.06$.

Larger deviations from the geometrical capacitance can be observed in the cleanest p-GaAs/GaAlAs heterojunction-insulated-gate
field-effect transistors (HIGFETs)~\cite{Huang}. In such devices a 2D hole gas is created with concentration as small as $n = 7 \times 10^8$ cm$^{-2}$ by a metallic gate at a distance $d$ from the 2D gas which can be as small as $250$ nm. This gives $nd^2 \sim 0.5$, so that if disorder permits $(d-d^*)/d \sim 0.2$ can be reached. There is no published data on capacitance for this case, but there are indications that screening of the Coulomb interaction between 2D holes by the gate (hole images) plays an important role for transport properties~\cite{Huang}.

One may be able to reach even larger $(d-d^*)/d$ for the capacitance
between two 2DEGs residing in two parallel quantum wells, because
in this case $d$ can be as small as $30$ nm~\cite{Eis1992,Eis-long}
so that $nd^2 < 1$ already at $n < 10^{11}$ cm$^{-2}$. We are not
aware of any such measurements.

Another system which may provide a good opportunity for studying capacitance larger than $C_g$ is that of a very low density 2DEG which floats on the top of a thin liquid helium film covering a metallic
electrode~\cite{Grimes,Konstantinov}.

Let us now turn to the most spectacular data, obtained from a
YBCO/LAO/STO capacitor~\cite{Ashoori}. We see from Fig.\
\ref{fig:Ashooricompare} that in this case a much smaller value $nd^2 \sim 0.02$ has been reached than in other cases, resulting
in the record for larger-than-geometrical capacitance
$(d-d^*)/d = 0.4$. This became possible because of the very small
distance between the 2DEG and the gate $d \simeq 4$ nm. Even at relatively large concentration $n = 10^{11} $cm$^{-2}$, such small thickness leads to $nd^2 \simeq 0.04$. In Fig.\ \ref{fig:Ashooricompare} the agreement of our theory (which has no adjustable parameters) with the experimental data of  Ref.~\cite{Ashoori} looks so good that disorder apparently plays a minor role.  The relatively large concentration of electrons may result in a significant resistance to disorder, but the relative unimportance of disorder is nonetheless difficult to understand. No independent estimates of disorder effects in YBCO/LAO/STO capacitor are currently
available in literature.

Our comparison with the data of Ref.~\cite{Ashoori} assumes that
the 2DEG is localized within a very narrow layer of STO, on the order of one lattice constant, as was shown in Refs.~\cite{Sing,Zhong}. There
are arguments that the electron layer is actually  much wider \cite{Siemons,Copie}, but these are based on calculations of the nonlinear screening radius of STO using its huge low-temperature
dielectric constant $\varepsilon \sim 20,000$. We disagree with using
the dielectric constant of STO as measured in uniform crystals when
describing nonlinear screening.  Indeed, the large dielectric
constant of STO (as well as that of displacement ferroelectric crystals) has a strong spatial dispersion. Below we discuss the origin of this dispersion following Ref.~\cite{Bursian}.

The large, low-temperature, zero-frequency dielectric constant of STO is related to the anomalously small frequency of the transverse optical
mode $\omega(q)$ at $q=0$. At $q=0$, the dielectric constant
$\varepsilon(0) \propto \omega^{-2}(0)$. For finite $q$, the dielectric
constant $\varepsilon(q) \propto \omega^{-2}(q)$. The soft mode is
known to have very strong dispersion, so that at large $q$ it
returns to the normal optical mode frequency. This dispersion has
the form $\varepsilon(q) = \varepsilon(0) /[1 + (q\xi)^2]$, where $\xi = a_0 \varepsilon^{1/2}(0)$ and $a_0$ is the lattice constant~\cite{Bursian}.  Using such a strongly dispersive dielectric constant for the description of nonlinear screening by electrons in STO self-consistently leads to the conclusion that the nonlinear screening radius, or in other words the width of the 2DEG, is of the order of the lattice constant $a_0$. (This same dispersion also explains why the large dielectric constant of STO does not lead to strong electron-phonon coupling and a large increase of the critical temperature of superconductivity~\cite{Meevasana}.)

\section{Conclusion} \label{sec:conclusion}

In this paper we have shown that in devices where a 2DEG comprises one or both electrodes of a plane capacitor the correlations between electronic charge in opposite electrodes dramatically affect the capacitance at low electron density $n$.  In the absence of disorder, this leads to a capacitance that grows strongly with decreasing $nd^2$, with a maximum value corresponding to $d^* = a/4$ at $nd^2 \rightarrow 0$.  We have presented a prediction for the effective capacitor thickness $d^*$ which is valid over the entire range of $nd^2$ and which is based on the Coulomb correlations between electrons and their image charges.  The cases of a 2DEG parallel to a metal electrode and of two parallel 2DEGs were considered separately.  Our results compare favorably to the recent experiments of Ref.\ \onlinecite{Ashoori}, which operate at $nd^2 \ll 1$ and report $C$ larger than $C_g$ by as much as 40\%, without the use of adjustable parameters.

The experiments of Ref.\ \onlinecite{Ashoori}, which use a 2DEG created at the an LAO/STO interface, are consistent with the 2DEG occupying a very thin layer of STO on the order of one lattice constant, so that the 2DEG can indeed be treated as a two-dimensional system up to fairly high electron density.  Further studies on such systems with low disorder and small LAO thickness may provide even better insight into the behavior of the capacitance at small $nd^2$.  Systems of very clean HIGFETs, parallel quantum wells, and electrons floating on liquid helium may also provide good opportunities for studying larger-than-geometrical capacitance.

\begin{small}
\vspace*{2ex} \par \noindent
{\em Acknowledgments.}

We are grateful to R.\ C.\ Ashoori, A.\ L.\ Efros, J.\ Eisenstein, M.\ M.\ Fogler, T.\ Kopp, L.\ Li, M.\ S.\ Loth, and B.\ Z.\ Spivak for helpful discussions.  B.\ S.\ acknowledges the support of the NSF Graduate Research Fellowship.

\end{small}

\end{document}